\documentclass[english,aps,reprint]{revtex4-1}
\usepackage[T1]{fontenc}
\usepackage[latin9]{inputenc}
\usepackage{amsmath}
\usepackage{amssymb}
\usepackage{graphicx}

\makeatletter

\newcommand*\LyXThinSpace{\,\hspace{0pt}}

\makeatother

\usepackage{babel}
\begin{document}
\title{Dirac Polarons and Resistivity Anomaly in $\mathrm{ZrTe}_{5}$ and
$\mathrm{HfTe}_{5}$}
\author{Bo Fu}
\affiliation{Department of Physics, The University of Hong Kong, Pokfulam Road,
Hong Kong, China}
\author{Huan-Wen Wang}
\affiliation{Department of Physics, The University of Hong Kong, Pokfulam Road,
Hong Kong, China}
\author{Shun-Qing Shen}
\email{sshen@hku.hk}

\affiliation{Department of Physics, The University of Hong Kong, Pokfulam Road,
Hong Kong, China}
\begin{abstract}
Resistivity anomaly, a sharp peak of resistivity at finite temperatures,
in the transition-metal pentatellurides $\mathrm{ZrTe_{5}}$ and $\mathrm{HfTe_{5}}$
was observed four decades ago, and more exotic and anomalous behaviors
of electric and thermoelectric transport were revealed recent years.
Here we present a theory of Dirac polarons, composed by massive Dirac
electrons and holes in an encircling cloud of lattice displacements
or phonons at finite temperatures. The chemical potential of Dirac
polarons sweeps the band gap of the topological band structure by
increasing the temperature, leading to the resistivity anomaly. Formation
of a nearly neutral state of Dirac polarons accounts for the anomalous
behaviors of the electric and thermoelectric resistivity around the
peak of resistivity.
\end{abstract}
\maketitle

\paragraph{Introduction}

Resistivity in the transition-metal pentatellurides $\mathrm{ZrTe}_{5}$
and $\mathrm{HfTe}_{5}$ exhibits a sharp peak at a finite temperature
$T_{p}$. The peak occurs approximately at a large range of temperatures
from 50 to 200K, but the exact value varies from sample to sample.
The effect was observed forty years ago \citep{okada1980giant,izumi1981anomalous},
but has yet to be understood very well. At the beginning, it was thought
as a structural phase transition, or occurrence of charge density
wave. The idea was soon negated as no substantial evidence is found
to support the picture \citep{disalvo1981possible,okada1982negative,bullett1982absence,fjellvag1986structural}.
The measurements of the Hall and Seebeck coefficients showed that
the type of charge carriers dominating the electrical transport changes
its sign around the peak, which indicates the chemical potential of
the charge carriers sweeps band gap around the transition temperature
$T_{p}$ \citep{izumi1982hall,jones1982thermoelectric,littleton1999transition,tritt1999enhancement}.
Thus the anomaly is believed to originate in the strong temperature
dependence of the chemical potential and carrier mobility. Recent
years the advent of topological insulators revives extensive interests
to explore the physical properties of $\mathrm{ZrTe}_{5}$ and $\mathrm{HfTe}_{5}$.
The first principles calculation suggested that the band structures
of $\mathrm{ZrTe}_{5}$ and $\mathrm{HfTe}_{5}$ are topologically
nontrivial in the layered plane or very close to the topological transition
points \citep{weng2014transition}. Further studies uncover more exotic
physics in these compounds \citep{chen2017spectroscopic,manzoni2016evidence,jiang2020unraveling,chen2015optical,li2016chiral,Zhou2016Pressure,Li2018Giant,Wang2018Discovery,liang2018anomalous,wang2019log,Zhang2019anomalous,tang2019three,Hu2019Large,wang2020quantum},
such as the chiral magnetic effect and three-dimensional quantum Hall
effect. Other possible causes have been advanced much recently \citep{Zhao2017anomalous,PhysRevX.8.021055,xu2018temperature},
but the physical origin of the resistivity anomaly is still unclear.
For example, it was suggested that a topological quantum phase transition
might occur, and the gap closing and reopening give rise to the resistivity
anomaly \citep{xu2018temperature}. However it contradicts with the
observation of the angle-resolved photoemission spectroscopy (ARPES)
measurement \citep{zhang2017electronic,zhang2017temperature}.

Strong temperature dependence of the band structure \citep{zhang2017electronic}
implies that the interaction between the Bloch electrons and the lattice
vibrations, \textit{i.e.}, electron-phonon interaction (EPI), is an
indispensable ingredient to understand the anomaly \citep{rubinstein1999hfte}.
In this Letter, we consider an anisotropic Dirac model describing
the low energy excitations of weak topological insulator near the
Fermi surface and EPI in $\mathrm{ZrTe_{5}}$ and $\mathrm{HfTe_{5}}$,
and propose a theory of Dirac polarons for the resistivity anomaly
at finite temperatures. The Dirac polarons are mixtures of massive
Dirac electrons and holes encircling a cloud of phonons, and are the
effective charge carriers in the compounds. Increasing temperature
will change the overlapping of the Dirac polarons drastically. The
chemical potential of Dirac polarons sweeping the band gap from conduction
bands to valence bands with increasing the temperature. Consequently,
when the chemical potential of Dirac polarons locates around the middle
of the band gap, the resistivity is enhanced drastically to form a
pronounced peak at a finite temperature. The carriers dominated the
charge transport change the sign around the transition. The formation
of a nearly neutral state of Dirac polarons accounts for anomalous
electric and magneto transport properties in the compounds.

\begin{figure}
\includegraphics[width=8.5cm]{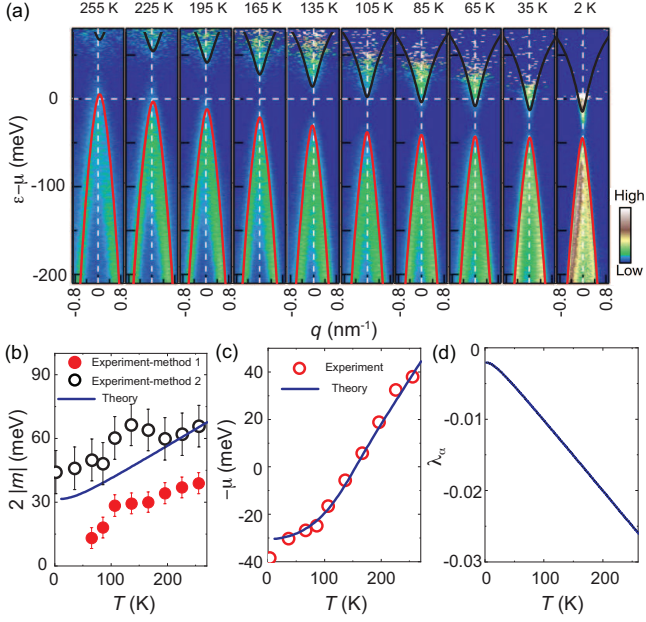}\caption{(a) A comparison of the renormalized energy spectrums according to
the theory (the solid lines) and the temperature-dependent band structures
from ARPES measurement adopted from Fig. 2(b) in Ref. \citep{zhang2017electronic}.
(b) The renormalized Dirac mass $m$ due to the EPI , (c) the temperature
dependent chemical potential $\mu$, and (d) the velocity dressing
function $\lambda_{\alpha}$. The experiment data are extracted from
Fig.2(d) and (f) in Ref. \citep{zhang2017electronic}. The model parameters
are set to be $v_{x}\simeq v_{y}=v_{\perp}=4\times10^{5}\mathrm{m}/s$,
$v_{z}=0.5\times10^{5}m/s$, $b_{x}\simeq b_{y}=b_{\perp}=230\mathrm{meV}\mathrm{nm}^{2}$,
$d_{x}\simeq d_{z}=d_{\perp}=-225\mathrm{meV}\mathrm{nm}^{2}$, $b_{z}=d_{z}=0$
and $m=12.0\,\mathrm{meV}$ for all the figures if there is no further
claiming. The carrier density used here is $n=4\times10^{17}\mathrm{cm^{-3}}$.}
\end{figure}

\paragraph{Finite temperature spectral function and quasiparticle properties}

The charge carriers in the conduction and valence bands of the bulk
$\mathrm{ZrTe_{5}}$ and $\mathrm{HfTe_{5}}$ are strongly coupled
together due to spin-orbit interaction and behave like massive Dirac
fermions instead of conventional electrons in semiconductors and metals
\citep{Wu2016evidence,Li2016Experimental,manzoni2016evidence}. In
the following, we only focus on $\mathrm{ZrTe_{5}}$ for comparison
with experimental measurement and theoretical calculation without
loss of generality. When the electrons (or holes) are moving through
the ionic lattices, the surrounding lattice will be displaced from
the original equilibrium positions; consequently, the electrons (or
holes) will be encircled by the lattice distortions, or phonons. At
finite temperatures, Dirac polarons are composed of both massive Dirac
electrons and holes in a cloud of phonons due to the thermal activation
when the chemical potential is located around the band edges as illustrated
in Fig.1(A). The Hamiltonian describing the EPI in Dirac materials
has the form \citep{Mahan-book}, $\mathcal{H}_{tot}=\mathcal{H}_{Dirac}+\mathcal{H}_{ph}+\mathcal{H}_{ep}$.
Here the phonon part $\mathcal{H}_{ph}$ is in the harmonic approximation,
and the EPI part $\mathcal{H}_{ep}$ is dominantly contributed by
longitudinal acoustic phonons. The low-energy physics of the electronic
states near the Fermi surface $\mathcal{H}_{Dirac}$, can be well
described by the anisotropic Dirac model \citep{SQS}, 
\begin{equation}
\mathcal{H}_{Dirac}(\mathbf{p})=(d(\mathbf{p})-\mu)I+\sum_{i=x,y,z}\hbar v_{i}p_{i}\alpha_{i}+m(\mathbf{p})\beta,\label{eq:DiracHamiltonian-1}
\end{equation}
where $\mathbf{p}=(p_{x},p_{y},p_{z})$ is the relative momentum to
the $\Gamma$ point, $v_{i}$ ($i=x,y,z$) are the effective velocities
in three directions. $\mu$ is the chemical potential. $d(\mathbf{p})=\sum_{i=x,y,z}d_{i}p_{i}^{2}$
breaks the particle-hole symmetry, and plays essential role in the
Dirac polaron physics. $m(\mathbf{p})=m-\sum_{i=x,y,z}b_{i}p_{i}^{2}$
is the momentum dependent Dirac mass. The first principles calculation
suggested that $\mathrm{ZrTe_{5}}$ is possibly a weak topological
insulator \citep{weng2014transition}, and the ARPES measurement showed
that there is no surface states within the band gap in its a-c plane
(the layers stacking along the b axis) \citep{weng2014transition}.
Thus, we consider an anisotropic case of $b_{x}\simeq b_{y}>0$ and
$b_{z}\leq0$. The detailed analysis of the band topology can be found
in Sec. SI of Ref. \citep{Note-on-SM} . The Dirac matrices are chosen
to be $\boldsymbol{\alpha}=\tau_{x}\otimes(\sigma_{x},\sigma_{y},\sigma_{z})$
and $\beta=\tau_{z}\otimes\sigma_{0}$, where $\sigma$ and $\tau$
are the Pauli matrices acting on spin and orbital space, respectively.
The quantitative information about these physical properties, such
as the Dirac velocity, the Dirac mass or the energy gap can be extracted
from the ARPES data \citep{Manzoni2015ultrafast,Moreschini2016nature,zhang2017electronic}.
To explore the EPI effect, we treat $\mathcal{H}_{ep}$ as a perturbation
to either electrons or phonons in the Migdal approximation \citep{Migdal-58jetp}
that the self-energy arises from the virtual exchange of a phonon
at temperature $T$. Due to the spinor nature of Dirac electrons,
the retarded self-energy can be recast in a matrix form as \citep{Note-on-SM,Garate2013phonon,Saha2014phonon}
\begin{equation}
\Sigma_{ep}^{R}(\mathbf{p},\epsilon)=\Sigma_{I}(\mathbf{p},\epsilon)+\lambda_{\alpha}(\mathbf{p},\epsilon)\hbar v\mathbf{p}\cdot\boldsymbol{\alpha}+\Sigma_{\beta}(\mathbf{p},\epsilon)\beta,
\end{equation}
where $\Sigma_{I}(\mathbf{p},\epsilon)$ is the renormalization to
the chemical potential $\mu$,$\lambda_{\alpha}(\mathbf{p},\epsilon)$
is the velocity dressing function and $\Sigma_{\beta}(\mathbf{p},\epsilon)$
is the renormalization to the Dirac mass $m$.

The quasiparticle properties of Dirac polaron can be obtained by the
poles of the retarded Green's function $G^{R}(\mathbf{p},\epsilon)=[\epsilon-\mathcal{H}_{Dirac}(\mathbf{p})-\Sigma^{R}(\mathbf{p},\epsilon)]^{-1}$,
which is in the complex plane with its real part gives the spectrum
of the quasiparticle and the imaginary part gives its lifetime. The
self-energy $\Sigma^{R}(\mathbf{p},\epsilon)=\Sigma_{ep}^{R}(\mathbf{p},\epsilon)+\Sigma_{imp}^{R}(\mathbf{p},\epsilon)$
includes the contribution from the impurities scattering. The spectral
function of the quasiparticle properties of Dirac polarons is given
by the imaginary part of $G^{R}(\mathbf{p},\epsilon)$, 
\begin{equation}
A_{\zeta}(\mathbf{p},\epsilon)=-\frac{1}{\pi}\langle\zeta s\mathbf{p}|\mathrm{Im}G^{R}(\mathbf{p},\epsilon)|\zeta s\mathbf{p}\rangle\text{,}
\end{equation}
where $|\zeta s\mathbf{p}\rangle$ are the band states with the band
indices $\zeta=\pm$ for the conduction and valence band and spin
indices $s=\pm$. In the absence of disorder and EPI, $A_{\zeta}(\mathbf{p},\epsilon)$
is a $\delta$ function reflecting that $\mathbf{p}$ is a good quantum
number and all its weight ratio is precisely at $\epsilon=\xi_{\mathbf{p}}^{\zeta}$.
In the presence of disorder and EPI, at low temperatures, $A_{\zeta}(\mathbf{p},\epsilon)$
exhibits a sharp peak of the Lorentzian type due to a long lifetime.
As temperature increases, $A_{\zeta}(\mathbf{p},\epsilon)$ maintains
the Lorentzian line shape but becomes broader due to the increasing
of the scattering rate, and the peak position moves to the positive
energy due to the renormalization of the energy level. The trajectories
of the peaks of the spectral function give us the renormalized dispersion
$\widetilde{\xi}_{\zeta}(\mathbf{p})$. As shown in Fig. 1(a), we
plot the derived energy dispersions $\widetilde{\xi}_{\zeta}(\mathbf{p})$
for different temperatures with the black and red lines corresponding
the conduction and valence band respectively. The ARPES data extracted
from Ref. \citep{zhang2017electronic} are also presented as the background
for a comparison. The excellent agreement can be found between our
theoretical calculations and the experiment data. The overall band
structure shifts up to higher energy with increasing temperature.
The peak structure of the spectral function can be clearly observed
in the temperature range considered, which suggests that a quasiparticle
picture is still appropriate at low energy and the EPI largely preserving
the weakly perturbed Fermi-liquid behavior.

The renormalized Dirac mass $m$ is given by the difference between
two energy levels $\widetilde{\xi}_{+}(\mathbf{0})$ and $\widetilde{\xi}_{-}(\mathbf{0})$
for the states at the band edge ($\mathbf{p}=0$):$\bar{m}=\frac{1}{2}[\widetilde{\xi}_{+}(\mathbf{0})-\widetilde{\xi}_{-}(\mathbf{0})]$.
At higher temperature, the effective mass varies with $T$ as $\bar{m}\simeq m+g_{m}T$
shown in Fig. 1(b). The coefficient $g_{m}$ are determined by the
band structure and the EPI strength (see the details in Ref.\citep{Note-on-SM}).
For Dirac materials, the renormalization of the energy levels is attributed
to the contributions from both intra- and inter-band scatterings.
With increasing the temperature, the more phonon modes with high momenta
are active, the larger the renormalization is. The chemical potential
is determined by the charge carriers density $n=\int_{|\bar{m}|}^{\infty}d\omega\left[\bar{\nu}_{+}(\omega)n_{F}(\omega-\mu)-\bar{\nu}_{-}(-\omega)n_{F}(\omega+\mu)\right]$
where $n_{F}(x)=[\exp(x/k_{B}T)+1]^{-1}$ is the Fermi distribution
function and $\bar{\nu}_{\pm}(\omega)$ are the renormalized density
of states for the conduction and valence band, respectively. In the
band structure of $\mathrm{ZrTe_{5}}$, the particle-hole symmetry
is broken and the valence band is narrower than the conduction band.
At the fixed $n$, the temperature dependence $\mu(T)$ are plotted
in Fig. 1(c). The calculated results demonstrate that the chemical
potential sweeps over the energy band gap of the massive Dirac particles
with increasing the temperature. At low temperatures, $\mu(T)\approx\mu(0)-\frac{\pi^{2}}{6}(k_{B}T)^{2}\frac{d\bar{\nu}_{\pm}(\omega)/d\omega}{\bar{\nu}_{\pm}(\omega)}\Big|_{\omega=\mu(0)}$shows
a quadratic temperature dependence by means of the Sommerfeld expansion.
$\mu(T)=0$ means the chemical potential is located at the mid-gap,
which approximately defines the transition temperature $T_{p}$ around.
At high temperatures, due to the strong particle-hole asymmetry and
the relatively low carrier density, the chemical potential shifts
into the valence band in a relatively linear fashion with increasing
the temperature. The velocity dressing function $\lambda_{\alpha}$
as a function of $T$ is plotted in Fig. 1(d). The velocity for Dirac
polaron decreases linearly with $T$ for higher temperature and saturates
a constant value for lower temperature.

\begin{figure}
\includegraphics[width=8.5cm]{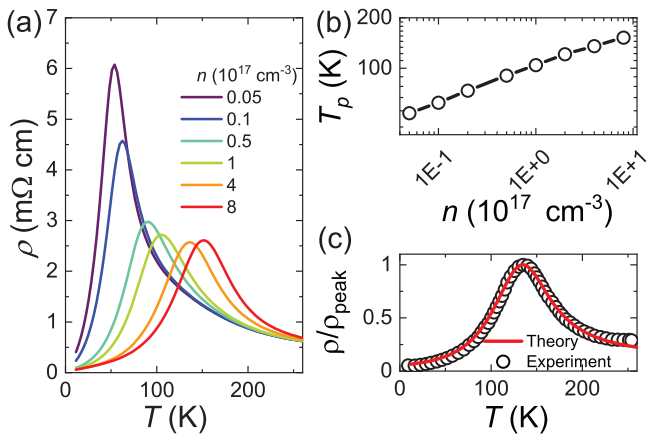}\caption{(a) The zero-field resistivity $\rho$ as a function of temperature
for several carrier density $n$. (b) The peak temperature $T_{p}$
as a function of the carrier density. (c) The comparison of the experimental
data and theoretical prediction by using the same parameters as Fig.
1. The experimental data are extracted from Fig. 1(d) in Ref. (\citep{zhang2017electronic}).
Both resistivity curves have been normalized to their maximum values
$\rho_{\mathrm{peak}}$.}
\end{figure}

\paragraph{The resistivity anomaly}

With the phonon-induced self-energy in hand, we are ready to present
the electrical resistivity as a function of temperature by means of
the linear response theory \citep{Mahan-book,Note-on-SM}. At finite
temperatures, the conductivities and thermoelectric coefficients are
contributed from both the electron-like and hole-like bands after
the phonon-induced renormalization. The two contributions are weighted
by the negative energy derivative of the Fermi-Dirac function, whose
value is nearly zero except for energies within a narrow window of
$k_{B}T$ near the chemical potential $\mu$. Figure 2(a) reproduces
the resistivity peak at several initial chemical potentials $\mu$,
or equivalently carrier densities at $T=0$. For the initial $\mu(T=0)$
($>0$) locating in the conduction band, as it moves down to the valance
band with increasing temperature, it will inevitably sweep over the
band gap. When $T=T_{p}$, the effective chemical potential lies around
the middle of the effective band gap $\mu(T=T_{p})\simeq0$ and the
resistivity reaches the maximum. As the $n$-type carrier concentration
is decreased, the resistivity peak will move to the lower temperature
with the higher magnitude. The peak temperature as a function of the
carrier density is plotted in Fig. 2(b). For a lower carrier concentration,
the chemical potential reaches the middle of the band gap with a lower
temperature. The height of the resistivity peak is determined by the
ratio $\bar{m}(T_{p})/(k_{B}T_{p})$. With increasing the ratio, the
peak height increases drastically, and becomes divergent if $\bar{m}(T_{p})\gg k_{B}T_{p}$.
It explains why in some experiments with extreme low carrier concentration
no resistivity peak is observed \citep{liang2018anomalous,mutch2019evidence},
which can be regarded as the situation of $T_{p}\sim0$. Thus the
sweeping chemical potential over the band gap of Dirac fermions gives
rise to the resistivity anomaly at finite temperatures. We use the
model parameters in Fig. 1 to calculate the resistivity, which is
in a good agreement with the experimental data as shown in Fig. 2(c).The
slight deviation at the high temperature might be caused by neglecting
the contributions from the optical modes of phonons.

\begin{figure}
\includegraphics[width=8.5cm]{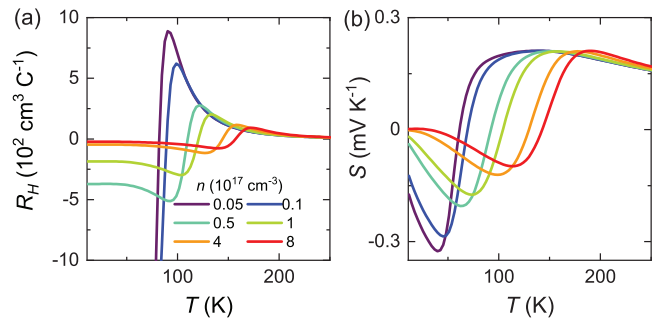}

\caption{(a) The Hall coefficient $R_{H}$ and (b) the Seebeck coefficient
$S$ as functions of temperature several carrier density $n$.}
\end{figure}

\paragraph{Sign change of the Hall and Seebeck coefficients}

The resistivity anomaly is always accompanied with the sign change
of the Hall and Seebeck coefficients around the transition temperature\citep{jones1982thermoelectric,chi2017lifshitz,zhang2020observation,tang2019three,Miller2018polycrystalline,Niemann2019magnetothermoelectric},
which can be reproduced in the present theory. As shown in Fig. 3(a),
for a positive $\mu(T=0)$ or $n$-type carriers , with increasing
the temperature, the Hall coefficient ($R_{H}=\partial\rho_{xy}/\partial B|_{B=0}$)
first maintains its value ($1/en$) at low temperature, decreases
down until reaching the minimum. Then $R_{H}$ changes from the negative
to positive sign at some temperature and continues to decrease down
to nearly zero at high temperatures. The sign change of $R_{H}$ indicates
the electron-dominated transport is transformed into the hole-dominated
as the chemical potential moves from the conduction band to valence
band. As the carrier concentration decreases, the Hall coefficient
crosses $0$ at a lower temperature with a larger maximum. In Fig.
3(b), the Seebeck coefficient $S_{xx}$ also reveals a systematic
shift in temperature as the carrier density increases. For each curve
with fixed carrier density, $S_{xx}$ displays similar nonmonotonic
temperature dependence as $R_{H}$, except that $S_{xx}$ starts from
absolute zero and exhibits a relative large positive ($p$-type) Seebeck
coefficient at high temperatures. At low temperatures, the chemical
potential lies deep in the bulk band, the Mott formula relates the
thermoelectric conductivity with the derivative of the electrical
conductivity for the thermopower $S_{xx}=\frac{\pi^{2}k_{B}^{2}T}{3e}\frac{d\sigma(\omega)/d\omega}{\sigma(\omega)}|_{\omega=\mu}$
\citep{Mott1969observation} with $\sigma(\omega)$ is the energy-dependent
conductivity. The conductivity $\sigma(\omega)$ is proportional to
the square of the group velocity. Hence, as chemical potential locates
in conduction band, $S_{xx}$ is negative ($n$-type) and decreases
with increasing temperature. $S_{xx}$ attains its largest value when
$n$ is tiny but nonvanishing, and varies rapidly with the temperature
around $T_{p}$. \citep{Nolas}. $T_{p}$ decreases with the reduction
of the $n$-type carrier concentration at zero temperature qualitatively
agrees with previous measurements for single crystals with different
carrier concentrations \citep{chi2017lifshitz}. Near $T=T_{p}$ and
if the band gap $\bar{m}(T_{p})$ is comparably smaller than the thermal
energy $k_{B}T_{p}$, either $R_{H}$ or $S_{xx}$ is linear in temperature
and the system enters a nearly neutral state of Dirac polarons due
to the strong thermal activation.

\begin{figure}
\includegraphics[width=8.5cm]{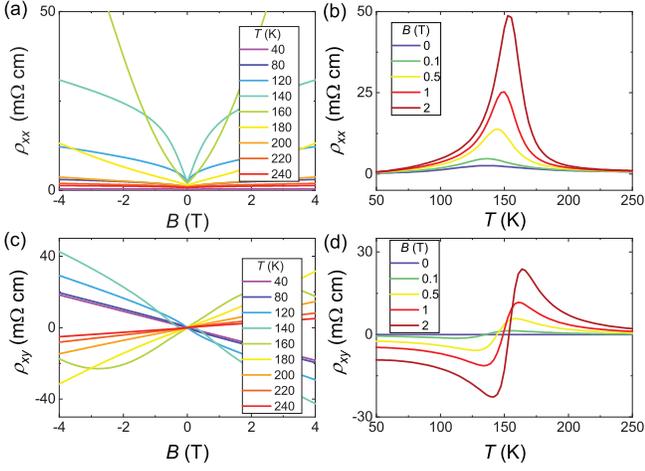}\caption{The magnetic field dependence of (a) the transverse magnetoresistance
$\rho_{xx}$ and (c) the Hall resistivity $\rho_{xy}$ for different
temperatures. The temperature dependence of (b) $\rho_{xx}$ and (d)
$\rho_{xy}$for different magnetic fields.}
\end{figure}

\paragraph{Magnetotransport in nearly neutral state of Dirac polarons}

The presence of an external magnetic field reveals the exotic behaviors
of magnetoresistivity near the transition temperature \citep{tritt1999enhancement,li2016chiral,tang2019three,Zhao2017anomalous,lv2018tunalbe,Niemann2019magnetothermoelectric}.
Without loss of generality we assume the magnetic field $B$is along
the $z$ direction. As shown in Fig. 4(a), the transverse magnetoresistivity
$\rho_{xx}(B)$ displays significantly different behaviors for temperature
above and below $T_{p}$. Below 120K, a narrow dip is observed around
zero magnetic field and above 200K, $\rho_{xx}$ shows a quadratic
field dependence. As approaching the peak temperature, $\rho_{xx}$
becomes large and nonsaturating. We plot the resistivity as a function
of temperature for different magnetic fields. As shown in Fig. 4(b),
$\rho_{xx}(B)$ displays striking resistivity peaks when the temperature
crosses the region of the neutral state of Dirac polarons. The peak
is strongly enhanced with increasing magnetic field, and even becomes
nonsaturated. Its position is observed to shift slightly to a higher
temperature with the field increasing, \textit{i.e.} $T_{p}$ is a
function of $B$. This effect has been reported experimentally in
Ref. \citep{li2016chiral,tang2019three}. The appearance of giant
and nonsaturated transverse magnetoresistivity can be viewed as the
electrical signature of the neutral state of Dirac polarons. As shown
in Fig. 4(c), the slope of the Hall resistivity $\rho_{xy}$ is negative,
indicating a electron-dominated charge transport. As the temperature
increases, the nonlinearity of $\rho_{xy}$ becomes more apparent.
In the intermediate temperature ($120\sim180$K) around $T_{p}$,
due to the formation of the nearly neutral state of Dirac polarons,
the slope of the Hall resistivity changes from positive (hole type)
at low magnetic field to negative (electron type) at high field, showing
a zigzag shaped profile. At high temperature (above 200K), the hole
carrier dominates the charge transport thus the slope of $\rho_{xy}$
become positive. The effect of an applied magnetic field on $\rho_{xy}$
as a function of temperature is shown in Fig. 4(d). There is a systematic
shift to the higher temperatures with increasing field. The calculated
$\rho_{xx}$ and $\rho_{xy}$ as functions of either $T$ or $B$
are in an excellent agreement with the experimental measurements in
$\text{ZrTe}_{5}$ and $\mathrm{HfTe}_{5}$ \citep{Zhao2017anomalous,tang2019three,Niemann2019magnetothermoelectric,lv2018tunalbe}.
Lastly, we want to point out the differences between the present theory
and the two-carrier model for magnetoresistance \citep{Pippard1989}.
The two-carrier model commonly requires that the Fermi surface is
composed of both electron and hole pockets and predicts a quadratical
magnetoresistance, while the present theory only involves a single
Dirac band crossing the Fermi surface and the multi-carrier transport
is attributted to thermal excitation over a wide range of temperature.

\paragraph{Discussion}

From an experimental standpoint, a temperature-dependent effective
carrier density can be deduced from the Hall measurement. The shift
of the chemical potential or effective carrier density with the variation
of temperature is the key issue to the resistivity anomaly. With no
absorption or desorption process through extrinsic doping, the temperature
dependent variation of effective density of charge carriers seems
to violate the conservation law of the total charge. However, the
relative contribution from each band of carriers to the total Hall
effect also depends on its ability to respond to the applied magnetic
field such as velocity and mobility. In Dirac materials with extreme
low carrier density and tiny band gap, the strong particle-hole asymmetry
will induce a significant temperature variation of the chemical potential,
even shifts from conduction band to valence band. Consequently, the
effective carrier density also displays strong temperature dependence.

We thank Li-Yuan Zhang, Nan-Lin Wang and Chen-Jie Wang for helpful
discussions. This work was supported by the Research Grants Council,
University Grants Committee, Hong Kong under Grant No. 17301717.

\newpage{}

\onecolumngrid

\section*{Supplementary Materials for ``Dirac Polarons and Resistivity Anomaly
in $\mathrm{ZrTe}_{5}$ and $\mathrm{HfTe}_{5}$''}

\section{The Model Hamiltonian for Anisotropic Dirac Materials}

For a three-dimensional quantum spin Hall system or topological insulator,
three are four $Z_{2}$ invariants to characterize 16 distinct phases,
in sharp contrast with two-dimensional case that only a single $Z_{2}$
topological invariant governs the effect \citep{fu2007topological-1}.
For a cubic lattice, three are 8 time reversal invariant momenta (TRIM)
expressed in terms of primitive reciprocal lattice vectors are $\Gamma_{i=(n_{1}n_{2}n_{3})}=(n_{1}\mathbf{b}_{1}+n_{2}\mathbf{b}_{2}+n_{3}\mathbf{b}_{3})/2$
with $n_{j}=0,1$. The four $Z_{2}$ topological invariants $(\nu_{0};\nu_{1}\nu_{2}\nu_{3})$
are defined as 
\begin{align*}
(-1)^{\nu_{0}} & =\prod_{n_{j}=0,1}\delta_{n_{1},n_{2},n_{3}},\\
(-1)^{\nu_{i=1,2,3}} & =\prod_{n_{j\ne i}=0,1;n_{i}=1}\delta_{n_{1},n_{2},n_{3}}.
\end{align*}
Generally, the calculation of $\delta_{i}$ requires a gauge in which
the wavefunctions are globally continuous, but in practice it is not
simple. When the system possesses the inversion symmetry, the problem
of identifying the $Z_{2}$ invariants is greatly simplified. In this
case, we only need to evaluate the expectation value of the parity
operator at the eight TRIMs which are the same for the two Kramers
degenerate states and 
\[
\delta_{i}=\prod_{m=1}^{N}\xi_{2m}(\Gamma_{i})
\]
where $\xi_{2m}(\Gamma_{i})=\pm1$ is the parity eigenvalue of the
degenerate states. The analytic expression for $\delta_{i}$ can be
obtained, 
\[
\delta_{i}=\frac{m(\Gamma_{i})}{|m(\Gamma_{i})|}.
\]

To identify the topology of the system, we need to take into account
the entire Brillouin zone, carefully examining the parity eigenvalues
at eight time-reversal invariant momenta. Our theory is based on $\boldsymbol{k}\cdot\boldsymbol{p}$
theory which is only valid around the Fermi level, NOT in the whole
Brillouin zone. In order to clarify the topology of the system, we
use the widely employed strategy to extend the low-energy continuous
model to a tight-binding model on a cubic lattice by replacing $k_{i}$
to $\sin k_{i}$ and $k_{i}^{2}$ to $2(1-\cos k_{i})$. Then the
mass term reads as $m(\mathbf{k})=m-2b_{x}(1-\cos k_{x})-2b_{y}(1-\cos k_{y})-2b_{z}(1-\cos k_{z})$.
Now we consider an anisotropic modified Dirac model with $b_{x}=b_{y}>0,b_{z}<0$
and $|b_{x}|>|b_{z}|$ because of the high anisotropic band structure
of $\mathrm{ZrTe}_{5}$ and $\mathrm{HfTe}_{5}$. For simplicity,
we assume it is nearly isotropic in the a-c plane. To ensure the lowest
energy electronic states for the entire spectrum is located at the
$\Gamma$ point, we can choose a small $|m|$. In this case, the four
$Z_{2}$ invariants can be obtained as,
\begin{align*}
(-1)^{\nu_{0}} & =sgn(m),\\
(-1)^{\nu_{1}} & =1,\\
(-1)^{\nu_{2}} & =1,\\
(-1)^{\nu_{3}} & =-1.
\end{align*}
Thus,the topological nature of this anisotropic Dirac model is controlled
by the sign of parameter $m$ such that an strong TI phase with $Z_{2}$
indices $(\nu_{0};\nu_{1}\nu_{2}\nu_{3})=(1,001)$ appears when $m<0$,
while the regime of $m>0$ falls into a weak TI phase with $(\nu_{0};\nu_{1}\nu_{2}\nu_{3})=(0,001)$.
This result is consistent with the first principles calculation in
Ref.(\citep{weng2014transition-1}).

The transport properties discussed in this work and its relevant physics
are determined mainly by the electrons near the Fermi surface. The
low-energy effective Dirac Hamiltonian successfully captures important
features of the band structure around the Fermi energy and the main
physics we are interested in. In the ARPES experiment, no surface
states can be observed within the gap in the the a-c plane (with the
layers stacking along the b axis) of $\mathrm{ZrTe}_{5}$ , with lowering
the temperature the energy gap tends to decrease and no topological
phase transition over the entire temperature range \citep{zhang2017electronic-1}.
To be consistent with the experimental observations, we adopt an anisotropic
Dirac model up to the quadratic term of the momentum which describes
the low-energy physics for the weak topological insulator
\[
\mathcal{H}_{Dirac}=(d(\mathbf{q})-\mu)I+\sum_{i=x,y,z}\hbar v_{i}p_{i}\alpha_{i}+m(\mathbf{p})\beta,
\]
with the particle-hole asymmetry term $d(\mathbf{q})=\sum_{i=x,y,z}d_{i}p_{i}^{2}$
and the mass term as $m(\mathbf{p})=m-\sum_{i=x,y,z}b_{i}p_{i}^{2}$.
For the convenience in explicit calculation, we use the parameters
as $v_{x}\simeq v_{y}=v_{\perp}$ $b_{x}\simeq b_{y}=b_{\perp}$,
$b_{z}=0$, $d_{x}\simeq d_{y}=d_{\perp}$and $d_{z}=0$, which can
be viewed as the minimal model for the band structure near the $\Gamma$
point of an anisotropic weak topological insulator.

\section{The Model for Electron-Phonon Interactions}

The Hamiltonian for lattice vibration in the harmonic approximation
can be expressed as
\begin{equation}
\mathcal{H}_{ph}=\sum_{\mathbf{q},\lambda}\hbar\omega_{\mathbf{q},\lambda}\left(a_{\mathbf{q}\lambda}^{\dagger}a_{\mathbf{q}\lambda}+\frac{1}{2}\right),\label{eq:barephononh-1}
\end{equation}
where $\omega_{\mathbf{q},\lambda}$ denotes the frequency of the
$\lambda$-th normal mode of wavevector $\mathbf{q}$. The DFT-calculations
demonstrate that the phonon energy ranges from $0$ to $28.9$ meV,
which corresponds an upper bound of the phonon frequency $\omega/2\pi\approx7\,\mathrm{THz}$
\citep{zhu2018record}. The velocities for the acoustic phonons at
$\Gamma$ point along $a,b,c$ axis are $2217$, $494$, and $2185\mathrm{ms^{-1}}$
\citep{zhu2018record}, respectively. The acoustic phonon velocity
along the $b$ axis is significantly smaller than those along the
$a,c$ axis. $\mathrm{Te}$ atoms give the dominate contribution to
the acoustic phonons as well as the low-energy optical modes $(<5.5\mathrm{THz})$
and due to the large stoichiometric ratio and heavier atomic mass.
Among total $36$ phonon bands ($12$ atoms in the primitive unite
cell), there are $3$ acoustic modes, $13$ inversion symmetry breaking
infrared-active optical modes, $18$ inversion symmetry preserving
Raman-active optical modes, and $2$ optical modes are optically inactive.
Further considering the constraint of the space-group symmetry of
$\mathrm{ZrTe}_{5}$, there are $6$ full crystalline symmetry protecting
$A_{g}$ Raman modes \citep{Aryal2020topological}. The variation
of the atomic displacement vectors for these modes will drive the
system into various topological phases. A Dirac topology surface separating
the strong topological insulator phase and the weak topological insulator
phase thus can be identified in the 6-dimensional space spanned by
these symmetry allowed Raman modes. The slight change in the lattice
parameters can be viewed as the superposition of the phonon modes
and corresponds to a single point in the formed multi-dimensional
space. Thus, the different sample growth conditions or some other
external perturbations such as strain and temperature may allocate
the system in distinct topological phases.

It is believed that acoustic phonons play a dominant role in modifying
the electronic properties and the carrier scattering at low temperature.
We consider that the EPI part $\mathcal{H}_{ep}$ is dominantly contributed
by longitudinal acoustic phonons, which can be expressed as
\begin{equation}
\mathcal{H}_{ep}=\sum_{\mathbf{q},\mathbf{k}}M_{\mathbf{q}}(a_{\mathbf{q}}+a_{-\mathbf{q}}^{\dagger})\psi_{\mathbf{k}+\mathbf{q}}^{\dagger}\psi_{\mathbf{k}}
\end{equation}
with the EPI strength as $M_{\mathbf{q}}=\sqrt{\hbar\mathbf{q}^{2}\Xi^{2}/(2V\varrho\omega_{\mathbf{q}})}$
where $\varrho=6.366\times10^{3}\mathrm{kgm^{-3}}$ is the atomic
mass density for $\mathrm{ZrTe_{5}}$, $\Xi=6\mathrm{eV}$ is the
acoustic deformation potential and $\omega_{\mathbf{q}}=c_{s}q$ is
the acoustic phonon frequency with the sound velocity chosen as $c_{s}=3040\,\mathrm{m}s^{-1}$,
$V=NV_{0}$ is the total volume with the unit cell volume for $\mathrm{ZrTe}_{5}$
as $V_{0}\approx400\mathring{\mathrm{A}}^{3}$ \citep{zhang2020observation}.
Note that electron-phonon scattering of the deformation potential
type conserves spin and pseudospin degrees of freedom. Here, for simplicity,
we adopt the isotropic model for $\mathrm{ZrTe}_{5}$. It is believed
that the anisotropy will only cause some quantitative, not quantitative
correction to the main results.

\section{The phonon-induced self-energy}

To explore the EPI effect, we treat $\mathcal{H}_{ep}$ as a perturbation
to either electrons or phonons. It will give rise to the quasiparticle
properties of the renormalized electrons and phonons. By definition,
the imaginary-time Green's functions for fermionic quasiparticle are
\begin{equation}
G(\mathbf{p},ip_{n})=-\int_{0}^{1/k_{B}T}d\tau e^{ip_{n}\tau}\langle T_{\tau}\psi_{\mathbf{p}}(\tau)\psi_{\mathbf{p}}^{\dagger}(0)\rangle
\end{equation}
and for bosonic quasiparticle
\begin{equation}
D(\mathbf{q},iq_{n})=-\int_{0}^{1/k_{B}T}d\tau e^{iq_{n}\tau}\langle T_{\tau}(a_{\mathbf{q}}(\tau)+a_{-\mathbf{q}}^{\dagger}(\tau))(a_{-\mathbf{q}}(0)+a_{\mathbf{q}}^{\dagger}(0))\rangle
\end{equation}
where $p_{n}=(2n+1)\pi k_{B}T$ and $q_{m}=2m\pi k_{B}T$ denote the
fermionic and bosonic Matsubara frequencies, respectively, with $n,m$
being integer numbers and $k_{B}$ is the Boltzmann constant. In the
interacting system, the electronic structure is characterized by the
renormalized finite-temperature Green's function $G(\mathbf{p},ip_{n})=[G^{(0)}(\mathbf{p},ip_{n})^{-1}+\Sigma(\mathbf{p},ip_{n})]^{-1}$
\citep{Mahan-book}. The bare Green's function for the unperturbed
Dirac Hamiltonian is
\begin{equation}
G^{(0)}(\mathbf{p},ip_{n})=\sum_{\zeta=\pm}\frac{\mathcal{P}_{\zeta}(\mathbf{p})}{ip_{n}-\xi_{\mathbf{p}}^{\zeta}},
\end{equation}
with the projection operators for the two bands are define as 
\begin{equation}
\mathcal{P}_{\zeta}(\mathbf{p})=\frac{1}{2}\left\{ 1+\left[\sum_{i}\hbar v_{i}p_{i}\alpha_{i}+m_{\mathbf{p}}\beta\right]]/\epsilon_{\mathbf{p}}^{\zeta}\right\} 
\end{equation}
and the eigenvalues $\xi_{\mathbf{p}}^{\zeta}=d(\mathbf{p})-\mu+\zeta\epsilon_{\mathbf{p}}$
with $\epsilon_{\mathbf{p}}=\sqrt{\sum_{i=x,y,z}\hbar^{2}v_{i}^{2}p_{i}^{2}+m^{2}(\mathbf{p})}$
are doubly degenerate for the conduction ($\zeta=+$) and valance
($\zeta=-$) bands which are measured with respect to the chemical
potential. In Dirac materials, the renormalization of the electron-phonon
vertex and higher order corrections to self-energy scale as the ratio
of sound to Fermi velocity $c_{s}/v_{F}$, which is a small quantity
in our problem. Thus, we only consider the lowest-order electron self-energy
arising from the virtual exchange of one phonon, which can be expressed
as \citep{Migdal-58jetp}

\begin{align}
\Sigma_{ep}(\mathbf{p},ip_{n}) & =-k_{B}T\sum_{iq_{m},\mathbf{q}}|M_{\mathbf{q}}|^{2}D^{(0)}(\mathbf{q},iq_{m})G^{(0)}(\mathbf{p}+\mathbf{q},ip_{n}+iq_{m})
\end{align}
with the bare phonon Green's functions $D^{(0)}(\mathbf{q},iq_{m})=\frac{-2\omega_{\mathbf{q}}}{q_{m}^{2}+\omega_{\mathbf{q}}^{2}}$.
After performing the Matsubara summation over frequencies $\omega_{n}$,
one obtains

\begin{equation}
\Sigma_{ep}(\mathbf{p},ip_{n})=\sum_{\zeta}\sum_{\mathbf{q}}|M_{\mathbf{q}}|^{2}\mathcal{P}_{\zeta}(\mathbf{p}+\mathbf{q})\left(\frac{n_{B}(\omega_{\mathbf{q}})+n_{F}(\xi_{\mathbf{p}+\mathbf{q}}^{\zeta})}{ip_{n}+\omega_{\mathbf{q}}-\xi_{\mathbf{p}+\mathbf{q}}^{\zeta}}+\frac{n_{B}(\omega_{\mathbf{q}})+1-n_{F}(\xi_{\mathbf{p}+\mathbf{q}}^{\zeta})}{ip_{n}-\omega_{\mathbf{q}}-\xi_{\mathbf{p}+\mathbf{q}}^{\zeta}}\right).\label{eq:self_energy-1}
\end{equation}
The self-energy depends on temperature via the Fermi-Dirac distribution
function $n_{F}(\varepsilon)=(e^{\varepsilon/k_{B}T}+1)^{-1}$and
Bose-Einstein distribution function $n_{B}(\varepsilon)=(e^{\varepsilon/k_{B}T}-1)^{-1}$,
respectively. Due to the EPI, for the state $|\zeta s\mathbf{p}\rangle$
with momentum $\mathbf{p}$ of the conduction band $\zeta=+$, it
can scatter virtually to the state $\xi_{\mathbf{p}+\mathbf{q}}^{\zeta}+\omega_{\mathbf{q}}$
with higher energy under simultaneous absorption of a phonon, or to
the state $\xi_{\mathbf{p}+\mathbf{q}}^{\zeta}-\omega_{\mathbf{q}}$
with lower energy under simultaneous emission of a phonon, respectively.
The integral over the momentum can be evaluated numerically by noticing
that the integrands depend only on the angle between momentum $\mathbf{p}$
and $\mathbf{q}$. To discuss the quasiparticle renormalization, the
retarded Green's function $G^{R}(\mathbf{p},\epsilon)$ and self-energy
$\Sigma^{R}(\mathbf{p},\epsilon)$ can be obtained by analytic continuation
to the real axis via $ip_{n}\to\epsilon+i\delta$ with an infinitesimal
positive $\delta$.

In some limiting regimes, the analytic expressions for these quantities
are available. At low temperatures $T\ll\Theta_{D}$, where $\Theta_{D}=\hbar c_{s}\Lambda/k_{B}$
is the Debye temperature with $\Lambda\sim\pi/a$ the high momentum
cutoff (in this work we use $\Lambda=6.28\mathrm{nm}^{-1}$ which
corresponds a Debye temperature $\Theta_{D}\sim150$K \citep{zhang2017electronic}),
the Bose-Einstein distribution function falls off exponentially for
$\hbar\omega_{\mathbf{q}}>k_{B}T$ and only the lower energy acoustic
phonons modes with long wavelength are active, the imaginary part
of self-energy follows a cubic temperature dependence and the real
part of self-energy saturate at a constant. Thus both the imaginary
and real part of the self-energy are linear in $T$. In high and low
temperature limits, the explicit expressions for imaginary part of
$\Sigma_{I}^{R}$,

\begin{align*}
\mathrm{Im}\Sigma_{\zeta}^{R}(\mathbf{p},\epsilon+i\delta) & =-\frac{\pi}{2}\sum_{\zeta^{\prime}}\sum_{\mathbf{q}}M_{\mathbf{q}}^{2}\mathrm{Tr}[\mathcal{P}_{\zeta}(\mathbf{p})\mathcal{P}_{\zeta^{\prime}}(\mathbf{p}+\mathbf{q})]\Bigg\{\left[n_{B}(\omega_{\mathbf{q}})+n_{F}(\xi_{\mathbf{p}+\mathbf{q}}^{\zeta^{\prime}})\right]\delta(\epsilon+\omega_{\mathbf{q}}-\xi_{\mathbf{p}+\mathbf{q}}^{\zeta^{\prime}})\\
 & +\left[n_{B}(\omega_{\mathbf{q}})+1-n_{F}(\xi_{\mathbf{p}+\mathbf{q}}^{\zeta^{\prime}})\right]\delta(\epsilon-\omega_{\mathbf{q}}-\xi_{\mathbf{p}+\mathbf{q}}^{\zeta^{\prime}})\Bigg\}
\end{align*}

Since the phonon energy is much smaller than the bulk band gap, the
delta function vanishes unless $\zeta=\zeta^{\prime}$. We further
set $\epsilon=\xi_{\mathbf{p}}^{\zeta}$, which is good approximation
when it is smaller than the chemical potential. For high temperature
$T\gg\Theta_{D}$, the Bose function takes the classical limit $n_{B}(\omega_{\mathbf{q}})\approx k_{B}T/(\hbar\omega_{\mathbf{q}})\gg1$,
the energy difference between two electron energies for the typical
momenta is much larger than the phonon energies $\omega_{\mathbf{q}}$
thus can be neglected in the delta functions. With these simplifications,
we have
\begin{align*}
\mathrm{Im}\Sigma_{\zeta}^{R}(\mathbf{p},\xi_{\mathbf{p}}^{\zeta}) & =-\frac{\pi}{2}\frac{\Xi^{2}k_{B}T}{\varrho c_{s}^{2}}\frac{1}{V}\sum_{\mathbf{p}^{\prime}}\mathrm{Tr}[\mathcal{P}_{\zeta}(\mathbf{p})\mathcal{P}_{\zeta}(\mathbf{p}^{\prime})]\delta(\xi_{\mathbf{p}}^{\zeta}-\xi_{\mathbf{p}^{\prime}}^{\zeta}).
\end{align*}
In this situation, the phonons can be viewed as the ``thermal static
disorder'' from the lattice, and the effective disorder strength
is proportional to $\Xi^{2}k_{B}T/(\varrho c_{s}^{2})$. By introducing
the density of states per band at the Fermi level $\nu(\mu)=\frac{1}{V}\sum_{\zeta}\sum_{\mathbf{p}}\delta(\xi_{\mathbf{p}}^{\zeta})$
and the Fermi-surface average of $A_{\mathbf{k}}$, 
\[
\langle A\rangle_{FS}=\left[\sum_{\xi}\sum_{\mathbf{k}}A_{\mathbf{k}}\delta(\xi_{\mathbf{p}}^{\zeta})\right]\Big/\left[\sum_{\xi}\sum_{\mathbf{k}}\delta(\xi_{\mathbf{p}}^{\zeta})\right],
\]
the imaginary part of the self-energy averaged over Fermi surface
at high temperature can be expressed as,
\[
\langle\mathrm{Im}\Sigma_{\zeta}^{R}(\mathbf{p},\xi_{\mathbf{p}}^{\zeta})\rangle_{FS}=-\frac{\pi}{2}\frac{\Xi^{2}k_{B}T}{\varrho c_{s}^{2}}\nu(\mu)\left[1+\langle\eta(\mathbf{p})\rangle_{FS}^{2}\right],
\]
with $\eta(\mathbf{p})=\frac{m_{\mathbf{p}}}{\epsilon_{\mathbf{p}}}$
the orbital polarization.

For low temperature $T\ll\Theta_{D}$, typical phonons have energy
$\omega_{\mathbf{q}}\sim k_{B}T$ and momenta $q\sim k_{B}T/\hbar c_{s}\ll\Theta_{D}/\hbar c_{s}\sim k_{F}$,
the argument of delta-function thus can be expanded as $\xi_{\mathbf{p}}^{\zeta}\pm\omega_{\mathbf{q}}-\xi_{\mathbf{p}+\mathbf{q}}^{\zeta}\approx\pm\omega_{\mathbf{q}}-\hbar\mathbf{v}_{\mathbf{p}}^{\zeta}\cdot\mathbf{q}$
with $\mathbf{v}_{\mathbf{p}}^{\zeta}=\frac{\partial\xi_{\mathbf{p}}^{\zeta}}{\hbar\partial\mathbf{p}}$
is the group velocity. After taking average over Fermi surface, we
have
\begin{align*}
\langle\mathrm{Im}\Sigma_{\zeta}^{R}(\mathbf{p},\xi_{\mathbf{p}}^{\zeta})\rangle_{FS} & \approx-\frac{\pi}{V}\sum_{\mathbf{p}}\sum_{\mathbf{q}}M_{\mathbf{q}}^{2}\left[n_{B}(\omega_{\mathbf{q}})+n_{F}(\omega_{\mathbf{q}})\right]\left[\delta(\omega_{\mathbf{q}}+\hbar\mathbf{v}_{\mathbf{p}}^{\zeta}\cdot\mathbf{q})+\delta(\omega_{\mathbf{q}}-\hbar\mathbf{v}_{\mathbf{p}}^{\zeta}\cdot\mathbf{q})\right]\delta(\xi_{\mathbf{p}}^{\zeta})\\
 & \approx-\pi\sum_{\mathbf{q}}M_{\mathbf{q}}^{2}\left[n_{B}(\omega_{\mathbf{q}})+n_{F}(\omega_{\mathbf{q}})\right]\left[\delta(\omega_{\mathbf{q}}+\hbar\langle|\mathbf{v}_{\mathbf{p}}^{\zeta}|\rangle_{FS}|\mathbf{q}|\cos\theta)+\delta(\omega_{\mathbf{q}}-\hbar\langle|\mathbf{v}_{\mathbf{p}}^{\zeta}|\rangle_{FS}|\mathbf{q}|\cos\theta)\right]\\
 & =-\frac{\hbar c_{s}\Xi^{2}}{4\pi\varrho c_{s}^{2}}\left(\frac{k_{B}T}{\hbar c_{s}}\right)^{3}\frac{1}{\hbar\langle|\mathbf{v}_{\mathbf{p}}^{\zeta}|\rangle_{FS}}\int_{0}^{\infty}dxx^{2}\left[\frac{1}{e^{x}+1}+\frac{1}{e^{x}-1}\right]\\
 & =-\frac{7\zeta(3)}{8\pi}\frac{\hbar\Xi^{2}}{\varrho c_{s}}\left(\frac{k_{B}T}{\hbar c_{s}}\right)^{3}\frac{1}{\hbar\langle|\mathbf{v}_{\mathbf{p}}^{\zeta}|\rangle_{FS}}.
\end{align*}
The final results for imaginary part of self-energy are collected
as,

\begin{align}
\langle\mathrm{Im}\Sigma_{\zeta}^{R}(\mathbf{p},\xi_{\mathbf{p}}^{\zeta})\rangle_{FS} & \approx\begin{cases}
-\frac{7\zeta(3)}{8\pi}\frac{\hbar\Xi^{2}}{\varrho c_{s}}\left(\frac{k_{B}T}{\hbar c_{s}}\right)^{3}\frac{1}{\hbar\langle|\mathbf{v}_{\mathbf{p}}^{\zeta}|\rangle_{FS}}, & T\ll\Theta_{D};\\
-\frac{\pi}{2}\frac{\hbar\Xi^{2}}{\varrho c_{s}}\frac{k_{B}T}{\hbar c_{s}}\nu_{\zeta}(\xi_{\mathbf{p}}^{\zeta})\left[1+\langle\eta(\mathbf{p})\rangle_{FS}^{2}\right], & T\gg\Theta_{D}.
\end{cases}
\end{align}

Now we calculate the real part of the self-energy. We concentrate
on the results at $\mathbf{p}=0$ which describe the energy renormalization
for the states at the band edge. Noticing that the dominant contribution
comes from the large momentum, it is a good approximation to further
let $\epsilon=0$. With these assumptions, the real part of self-energies
for the chemical potential, the Dirac mass and the velocity can be
calculated through,
\begin{align*}
\mathrm{Re}\Sigma_{I}(\mathbf{0},0) & =\frac{1}{2}\sum_{\chi}\sum_{\mathbf{q}}|M_{\mathbf{q}}|^{2}\left(\frac{n_{B}(\omega_{\mathbf{q}})+n_{F}(\xi_{\mathbf{q}}^{\chi})}{\omega_{\mathbf{q}}-\xi_{\mathbf{q}}^{\chi}}+\frac{n_{B}(\omega_{\mathbf{q}})+1-n_{F}(\xi_{\mathbf{q}}^{\chi})}{-\omega_{\mathbf{q}}-\xi_{\mathbf{q}}^{\chi}}\right),\\
\mathrm{Re}\Sigma_{\beta}(\mathbf{0},0) & =\frac{1}{2}\sum_{\chi}\sum_{\mathbf{q}}|M_{\mathbf{q}}|^{2}\chi\eta(\mathbf{q})\left(\frac{n_{B}(\omega_{\mathbf{q}})+n_{F}(\xi_{\mathbf{q}}^{\chi})}{\omega_{\mathbf{q}}-\xi_{\mathbf{q}}^{\chi}}+\frac{n_{B}(\omega_{\mathbf{q}})+1-n_{F}(\xi_{\mathbf{q}}^{\chi})}{-\omega_{\mathbf{q}}-\xi_{\mathbf{q}}^{\chi}}\right),\\
\lambda_{\alpha} & =\frac{1}{2}\sum_{\chi}\sum_{\mathbf{q}}|M_{\mathbf{q}}|^{2}\frac{1}{\epsilon_{\mathbf{q}}^{\chi}}\left(\frac{n_{B}(\omega_{\mathbf{q}})+n_{F}(\xi_{\mathbf{q}}^{\chi})}{\omega_{\mathbf{q}}-\xi_{\mathbf{q}}^{\chi}}+\frac{n_{B}(\omega_{\mathbf{q}})+1-n_{F}(\xi_{\mathbf{q}}^{\chi})}{-\omega_{\mathbf{q}}-\xi_{\mathbf{q}}^{\chi}}\right).
\end{align*}

In evaluating the real part of self-energy, the phonon energy $\omega_{\mathbf{q}}$
in the denominator is less important and can be neglected. The two
terms in the parentheses has the same denominator, then the Fermi
factors cancel when they are added. At low temperature, $n_{B}(\omega_{\mathbf{q}})\approx0$,
and at high temperature, $n_{B}(\omega_{\mathbf{q}})\approx k_{B}T/\omega_{\mathbf{q}}$.
After introducing some constants independent of temperature, 
\begin{align*}
\mathcal{C}_{\beta}^{H} & =\frac{\hbar c_{s}\Lambda}{\Lambda^{3}}\int_{|\mathbf{q}|<\Lambda}\frac{d^{3}\mathbf{q}}{(2\pi)^{3}}\frac{m(\mathbf{q})}{(d(\mathbf{q})-\mu)^{2}-\epsilon_{\mathbf{q}}^{2}};\mathcal{C}_{\beta}^{L}=\frac{1}{2\Lambda^{3}}\int_{|\mathbf{q}|<\Lambda}\frac{d^{3}\mathbf{q}}{(2\pi)^{3}}\frac{\hbar c_{s}qm(\mathbf{q})}{(d(\mathbf{q})-\mu)^{2}-\epsilon_{\mathbf{q}}^{2}};\\
\mathcal{C}_{I}^{H} & =\frac{\hbar c_{s}\Lambda}{\Lambda^{3}}\int_{|\mathbf{q}|<\Lambda}\frac{d^{3}\mathbf{q}}{(2\pi)^{3}}\frac{\mu-d(\mathbf{q})}{(d(\mathbf{q})-\mu)^{2}-\epsilon_{\mathbf{q}}^{2}};\mathcal{C}_{I}^{L}=\frac{1}{2\Lambda^{3}}\int_{|\mathbf{q}|<\Lambda}\frac{d^{3}\mathbf{q}}{(2\pi)^{3}}\frac{\hbar c_{s}q(\mu-d(\mathbf{q}))}{(d(\mathbf{q})-\mu)^{2}-\epsilon_{\mathbf{q}}^{2}};\\
\mathcal{C}_{\alpha}^{H} & =\frac{\hbar c_{s}\Lambda}{\Lambda^{3}}\int_{|\mathbf{q}|<\Lambda}\frac{d^{3}\mathbf{q}}{(2\pi)^{3}}\frac{1}{(d(\mathbf{q})-\mu)^{2}-\epsilon_{\mathbf{q}}^{2}};\mathcal{C}_{\alpha}^{L}=\frac{1}{2\Lambda^{3}}\int_{|\mathbf{q}|<\Lambda}\frac{d^{3}\mathbf{q}}{(2\pi)^{3}}\frac{\hbar c_{s}q}{(d(\mathbf{q})-\mu)^{2}-\epsilon_{\mathbf{q}}^{2}}.
\end{align*}
The real part of $\Sigma_{I}^{R}$ and $\Sigma_{\beta}^{R}$ can be
expressed as

\begin{align}
\mathrm{Re}\Sigma_{I,\beta}^{R}(\mathbf{0},0)=\frac{\Xi^{2}\Lambda^{3}}{\varrho c_{s}^{2}}\begin{cases}
\mathcal{C}_{I,\beta}^{L}, & T\ll\Theta_{D};\\
\frac{k_{B}T}{\hbar c_{s}\Lambda}\mathcal{C}_{I,\beta}^{H}, & T\gg\Theta_{D}.
\end{cases}\label{eq:massrenormalization}
\end{align}
Thus the effective Dirac mass due to EPI can be obtained as $\bar{m}=m+\mathrm{Re}\Sigma_{\beta}(\mathbf{0},0)$,
which exhibits strong temperature dependence. Since the EPI strength
is momentum-dependent, the velocity renormalization $\lambda_{\alpha}(\mathbf{p},\epsilon)$
do not vanish. In the high and low temperature limits, the renormalization
factor can be obtained as,

\[
\lambda_{\alpha}=\frac{\Xi^{2}\Lambda^{3}}{\varrho c_{s}^{2}}\begin{cases}
\mathcal{C}_{\alpha}^{L}, & T\ll\Theta_{D}\\
\frac{k_{B}T}{\hbar c_{s}\Lambda}\mathcal{C}_{\alpha}^{H}, & T\gg\Theta_{D}
\end{cases}
\]
which is negative and its absolute value gets larger with increasing
temperature. When electrons are dressed by a cloud of phonons, the
velocity is effectively reduced by the EPI as the temperature increases.
It is worth noting that this value differs from the noninteracting
case even at $T=0\mathrm{K}$ due to the zero-point vibration.

By definition, the density of the charge carriers is
\begin{equation}
n=\int_{|\bar{m}|}^{\infty}d\omega\bar{\nu}(\omega)n_{F}[\omega-\mu]-\int_{-\infty}^{-|\bar{m}|}d\omega\bar{\nu}(\omega)[1-n_{F}(\omega-\mu)],
\end{equation}
where $n_{F}(\omega-\mu)$ is the Fermi distribution function, $\bar{\nu}$
are the renormalized density of states for conduction and valence
band after considering the EPI. The temperature dependent chemical
potential can be determined by solving this equation with fixed total
number of carriers $n$.

As shown in Fig. 1(b) and (c), the calculated results for $\mu$ and
$\bar{m}$ demonstrate that the chemical potential shifts with temperature,
and sweeps over the energy band gap of the massive Dirac particles
for a proper choice of the model parameters. $\mu=0$ means the chemical
potential is located at the mid-gap, which approximately defines the
transition temperature $T_{p}$ around. At high temperature, these
two quantities follow linear temperature dependence: $\mu\approx g_{\mu}(T_{p}-T)$
and $\bar{m}=m+g_{m}T$ with the coefficient $g_{m}$ can be determined
from Eq. (\ref{eq:massrenormalization}),
\begin{align}
g_{m} & =\frac{\Xi^{2}\Lambda^{3}}{\varrho c_{s}^{2}}\frac{k_{B}}{\hbar c_{s}\Lambda}\mathcal{C}_{\beta}^{H},
\end{align}
and the coefficient $g_{\mu}$ can be obtained by fitting the numerical
results for $\mu$.

Furthermore we also take into account a disorder potential $U(\mathbf{r})$
to simulate the impurities which are distributed randomly in the sample.
We assume the disorder potential behaves like a white noise with the
correlator as, $\langle U(\mathbf{r})U(\mathbf{r}^{\prime})\rangle=n_{imp}U_{imp}^{2}$
which is chosen to be $n_{imp}U_{imp}^{2}=100\mathrm{meV}^{2}\mathrm{nm}^{3}$
for all the calculations. The static disorder will induce a self-energy,
\begin{equation}
\mathrm{Im}\Sigma_{imp}^{R}(\mathbf{p},\xi_{\mathbf{p}}^{\zeta})=-\frac{\pi}{2}n_{imp}U_{imp}^{2}\nu_{\zeta}(\xi_{\mathbf{p}}^{\zeta})\left[1_{4}+\zeta\eta(\mathbf{p})\beta\right].
\end{equation}
The real parts of the self-energy induced by disorder effect are expected
to be quite small in the weak scattering limit. We ignore this contribution
in the calculations.

\section{The renormalization of the energy level of Dirac polaron}

From Eq. (\ref{eq:self_energy-1}), the renormalization to the specific
electronic state due to EPI can be expressed as,
\begin{equation}
\widetilde{\xi}_{\mathbf{p}}^{\zeta}\approx\xi_{\mathbf{p}}^{\zeta}+\sum_{\zeta^{\prime}\mathbf{q}}|M_{\mathbf{q}}|^{2}\left(\sum_{s^{\prime}}|\langle\zeta s\mathbf{p}|\zeta^{\prime}s^{\prime}\mathbf{p}+\mathbf{q}\rangle|^{2}\right)\frac{2n_{B}(\omega_{\mathbf{q}})+1}{\xi_{\mathbf{p}}^{\zeta}-\xi_{\mathbf{p}+\mathbf{q}}^{\zeta^{\prime}}},\label{eq:energy_renormalization}
\end{equation}
where we have neglected the phonon frequencies in the denominator
since the electronic energy scale is much larger than the phonon energies
$(c_{s}\ll v)$. In conventional large band gap semiconductors, the
energy spacing between two near bands is large enough such that the
renormalization only comes from intraband scattering. However, for
Dirac materials, the energy denominators for interband scattering
are not so large that this part of contribution can not be ruled out,
the competition between intraband and interband contribution will
lead to rich physical results. In common sense, the large momentum
transfer processes are negligible due to the large energy denominator.
However, for Dirac spectrum, the electron-phonon scattering matrix
elements $M_{\mathbf{q}}\langle\zeta s\mathbf{p}|\zeta^{\prime}s^{\prime}\mathbf{p}+\mathbf{q}\rangle$
is getting larger as the momentum increases in the case of coupling
to longitudinal acoustic phonons and the phase space for higher energy
is larger. Therefore, the large momentum transfer processes plays
important roles in the formation of Dirac polarons.

The renormalizations of band gap and chemical potential can be obtained
from the energy difference $\delta m=\frac{1}{2}(\widetilde{\xi}_{\mathbf{0}}^{+}-\widetilde{\xi}_{\mathbf{0}}^{-})-m$
and the energy average $\delta\mu=-\frac{1}{2}(\widetilde{\xi}_{\mathbf{0}}^{+}+\widetilde{\xi}_{\mathbf{0}}^{-})$
of the two states at the band edge ($\mathbf{p}=0$). In this simplified
situation, the amplitudes of electron-phonon scattering matrix elements
for intraband and interband can be expressed as $\sum_{s^{\prime}}|M_{\mathbf{q}}\langle\zeta s\mathbf{0}|\zeta s^{\prime}\mathbf{q}\rangle|^{2}=|M_{\mathbf{q}}|^{2}\frac{1}{2}[1+\eta(\mathbf{q})]$
and $\sum_{s^{\prime}}|M_{\mathbf{q}}\langle\zeta s\mathbf{0}|\bar{\zeta}s^{\prime}\mathbf{q}\rangle|^{2}=|M_{\mathbf{q}}|^{2}\frac{1}{2}[1-\eta(\mathbf{q})]$,
respectively. Considering the $\mathbf{q}$ summation in Eq. (\ref{eq:energy_renormalization})
is dominated by the large momentum-transfer processes, we can neglect
the chemical potential and Dirac mass in the denominators, and finally
obtain the chemical potential renormalization 
\begin{equation}
\delta\mu=\frac{1}{2}\sum_{\mathbf{q}}\coth\frac{\omega_{\mathbf{q}}}{2k_{B}T}|M_{\mathbf{q}}|^{2}\left(\frac{1}{\xi_{\mathbf{q}}^{+}}+\frac{1}{\xi_{\mathbf{q}}^{-}}\right)
\end{equation}
and the Dirac mass renormalization 
\begin{equation}
\delta m=-\frac{1}{2}\sum_{\mathbf{q}}\coth\frac{\omega_{\mathbf{q}}}{2k_{B}T}|M_{\mathbf{q}}|^{2}\eta(\mathbf{q})\left(\frac{1}{\xi_{\mathbf{q}}^{+}}-\frac{1}{\xi_{\mathbf{q}}^{-}}\right).
\end{equation}
 When the particle-hole symmetry is preserved, the contributions from
conduction band and valance band compensate, $\delta\mu=0$ and the
chemical potential exhibits no shift. In contrast, when the particle-hole
symmetry is broken, for $\xi_{\mathbf{q}}^{+}>-\xi_{\mathbf{q}}^{-}$,
that the conduction band is narrower than the valance band, the chemical
potential is pushed down ($\delta\mu<0$) while for $\xi_{\mathbf{q}}^{+}<-\xi_{\mathbf{q}}^{-}$
the chemical potential is pulled up. As temperature increases, more
phonon modes with high momentum are active, the renormalization becomes
larger.
\begin{figure}
\includegraphics[width=10cm]{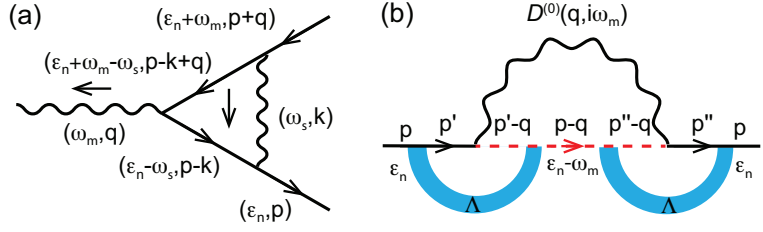}

\caption{(a) First order correction of the electron-phonon vertex beyond Migdal's
approximation. The wavy lines represent phonons and the solid lines
represent the fermions. (b) The electron-phonon interaction induced
self-energies dressed by Diffuson.}
\end{figure}

\section{The vertex correction beyond Migdal's approximation}

In this section, we will show that the vertex correction has the order
of $c_{s}/v_{F}$ which is relatively small and can be neglected in
the present work. According to the rules for the Feynman diagrams
depicted as Fig. S1(a), the first order perturbation vertex correction
beyond the Migdal's approximation can be expressed as 
\[
\Gamma(p,q)=\frac{1}{\beta V}\sum_{\omega_{s}}\sum_{\mathbf{k}}M_{\mathbf{k}}^{2}G^{(0)}(\epsilon_{n}-\omega_{s},\mathbf{p}-\mathbf{k})G(\epsilon_{n}-\omega_{s}+\omega_{m},\mathbf{p}-\mathbf{k}+\mathbf{q})D(\omega_{s},\mathbf{k}),
\]
where $D^{(0)}$ and $G^{(0)}$ are the electron and phonon Green's
function in the absence of interaction. We consider the external variables
$\epsilon_{n}=0$ and $\mathbf{p}=0$, which is the crucial case for
our further calculations, 
\begin{equation}
\Gamma(\omega_{m},q)=\sum_{\zeta,\zeta^{\prime}}\sum_{\omega_{s}}\sum_{\mathbf{k}}M_{\mathbf{k}}^{2}\frac{-2\omega_{\mathbf{k}}}{\omega_{s}^{2}+\omega_{\mathbf{k}}^{2}}\frac{\mathcal{P}_{\zeta}(\mathbf{k})}{i\omega_{s}-\xi_{\mathbf{k}}^{\zeta}}\frac{\mathcal{P}_{\zeta^{\prime}}(\mathbf{q}-\mathbf{k})}{i(\omega_{s}-\omega_{m})-\xi_{\mathbf{k}-\mathbf{q}}^{\zeta^{\prime}}}.\label{eq:vertexfunction}
\end{equation}
Notice that the dominant contribution in Eq. (\ref{eq:vertexfunction})
arises from the region of $\mathbf{k}$ space where the integrand
has a vanishing demonimator. Thus, we restrict our consideration to
the states crossing the Fermi energy $\zeta=\zeta^{\prime}=+$, such
that we can drop the band indices. By projecting onto the identity
matrix, substituting the electron-phonon coupling strength $M_{\mathbf{k}}^{2}=\frac{\hbar c_{s}|\mathbf{k}|\Xi^{2}}{2V\varrho c_{s}^{2}}$
and taking zero-temperature limit $T\sum_{\omega}\to\int_{-\infty}^{\infty}\frac{d\omega}{2\pi}$
, we have

\[
\Gamma(\omega_{m},q)=\frac{\gamma}{2\pi}\int_{-E_{c}}^{E_{c}}dE\int_{-\infty}^{\infty}d\omega\frac{1}{2}\int_{-1}^{1}dx\frac{\omega_{0}}{i\omega-E}\frac{-2\omega_{0}}{\omega^{2}+\omega_{0}^{2}}\frac{1}{i(\omega-\omega_{m})-(E-v_{F}qx)}.
\]
Here we have truncated to the linear order of the momentum and consider
a locally linear dispersion of the band in the vicinity of $k_{F}$,
$\xi_{\mathbf{k}}=\epsilon_{\mathbf{k}}-\mu\approx v_{F}|\mathbf{k}|-\mu\equiv E$
and $\xi_{\mathbf{k}-\mathbf{q}}=\epsilon_{\mathbf{k}-\mathbf{q}}-\mu\approx v_{F}|\mathbf{k}-\mathbf{q}|-\mu\approx E-v_{F}|\mathbf{q}|\cos\theta$
with $\theta$ is the scattering angle and introduced new integration
variables $E=v_{F}|\mathbf{k}|-\mu$ and $x=\cos\theta$. We choose
an energy cutoff of the order of Fermi energy $E_{c}\sim\hbar v_{F}k_{F}$.
Since we only work to the orders of magnitude, the density of state
$\nu$ is assumed to be constant between $-E_{c}$ and $E_{c}$ and
the phonon's energy is approximated as $\omega_{\mathbf{k}}\approx\hbar c_{s}k_{F}\equiv\omega_{0}$.
$\gamma=\frac{\Xi^{2}\nu}{2\varrho c_{s}^{2}}$ is the dimensionless
coupling strength. Then, we first evaluate the integration over $\omega$,
it is convenient to split the integrand into sum of relatively simple
terms:
\begin{align*}
\Gamma(\omega_{m},q) & =-\gamma\int_{-E_{c}}^{E_{c}}dx\frac{1}{2}\int_{-1}^{1}dx\frac{\omega_{0}}{\omega_{m}+iv_{F}qx}\int_{-\infty}^{\infty}d\omega\frac{1}{2\pi i}\Bigg[\frac{1}{(\omega-\omega_{m})+i(E-v_{F}qx)}\frac{1}{\omega+i\omega_{0}}-\frac{1}{\omega+ix}\frac{1}{\omega+i\omega_{0}}\\
 & -\frac{1}{(\omega-\omega_{m})+i(E-v_{F}qx)}\frac{1}{\omega-i\omega_{0}}+\frac{1}{\omega+iE}\frac{1}{\omega-i\omega_{0}}\Bigg].
\end{align*}
Only when the singularities of two energy denominators are on the
opposite side of the real axis, the integration will not vanish, which
leads to 
\begin{align*}
\Gamma(\omega_{m},q) & =-\gamma\int_{-E_{c}}^{E_{c}}dE\frac{1}{2}\int_{-1}^{1}dx\frac{\omega_{0}}{\omega_{m}+iv_{F}qx}\Bigg[\frac{\theta(-E+v_{F}qx)}{\omega_{m}-i(E-v_{F}qx)+i\omega_{0}}-\frac{\theta(-E)}{-iE+i\omega_{0}}\\
 & -\frac{\theta(E-v_{F}qx)}{i\omega_{0}-\omega_{m}+i(E-v_{F}qx)}+\frac{\theta(E)}{i\omega_{0}+iE}\Bigg].
\end{align*}
where $\theta(x)$ is the Heaviside step function. The second and
the forth terms in the bracket cancel each other when integrating
over $E$. Performing $E$ integration, the other two terms gives,
\begin{align*}
\Gamma(\omega_{m},q) & =-\gamma\int_{-1}^{1}dx\frac{1}{(v_{F}qx)^{2}+\omega_{m}^{2}}\Bigg\{\frac{1}{2}v_{F}qx\log\left(\frac{\omega_{m}^{2}+(E_{c}+\omega_{0}-v_{F}qx)^{2}}{\omega_{m}^{2}+\omega_{0}^{2}}\right)\\
 & -\omega_{m}\left[\tan^{-1}\left(\frac{\omega_{m}}{E_{c}+\omega_{0}-v_{F}qx}\right)-\tan^{-1}\left(\frac{\omega_{m}}{\omega_{0}}\right)\right]\Bigg\}.
\end{align*}
In order to proceed with the $x$ integration we further expand the
integrand for small value of $q$ up to quadratic terms. After the
integration, we arrive at the expression for vertex function,

\begin{align*}
\Gamma(\omega_{m},q) & =-2\gamma\frac{\omega_{0}}{v_{F}q}\Bigg\{\left[\tan^{-1}\left(\frac{\omega_{m}}{\omega_{0}}\right)-\tan^{-1}\left(\frac{\omega_{m}}{E_{c}+\omega_{0}}\right)\right]\tan^{-1}\frac{v_{F}q}{\omega_{m}}\\
 & -\frac{\omega_{m}(E_{c}+\omega_{0})}{(E_{c}+\omega_{0})^{2}+\omega_{m}^{2}}\left[1+\frac{\omega_{m}^{2}}{(E_{c}+\omega_{0})^{2}+\omega_{m}^{2}}\right]\left(\frac{v_{F}q}{\omega_{m}}+\tan^{-1}\frac{v_{F}q}{\omega_{m}}\right)\Bigg\}.
\end{align*}
Now we consider the static limit by taking $\omega_{m}\to0$ first
and then $q\to0$, 
\[
\lim_{q\to0}\lim_{\omega_{m}\to0}\Gamma(\omega_{m},q)=-\gamma\frac{2\omega_{0}}{E_{c}+\omega_{0}}=-2\gamma\frac{c_{s}}{v_{F}+c_{s}}
\]
When the velocity of sound $(c_{s})$ is much smaller than the Fermi
velocity $(v_{F})$ which is typically the situation in solid state
materials, the correction to the electron-phonon vertex is suppressed
by a factor $c_{s}/v_{F}$ and can be safely neglected.

\section{The vertex corrections to the electron-phonon self-energy from disorder
effect}

In the theory of disordered noninteraction system, the disordered
averaged product of Green's functions in the particle-hole polarization
bubble gives the ``diffusion'' mode at low frequencies and momenta
($|\omega|,v_{F}q\ll\tau_{0}^{-1}$, the so-called diffusion approximation,
with $v_{F}$ is the Fermi velocity). The diffusion mode will introduce
a vertex correction to the electron-phonon vertices,
\[
M_{\mathrm{diff}}(\mathbf{q},i\omega_{n})=M_{\mathbf{q}}\Lambda(\mathbf{q},i\omega_{n}).
\]
The particle-hole diffusion vertex $\Lambda(\mathbf{q},i\omega_{n})$
can be calculated through the summation of the ladder diagrams,
\[
\Lambda(\mathbf{q},i\omega_{n})=[1-P(\mathbf{q},i\omega_{n})]^{-1},
\]
with

\begin{align*}
P(\mathbf{q},i\omega_{n}) & =\frac{1}{2\pi\nu\tau_{0}}\int\frac{d^{3}\mathbf{k}}{(2\pi)^{3}}\mathrm{Tr}[G(\mathbf{k}-\mathbf{q},i\epsilon_{m}-i\omega_{n})G(\mathbf{k},i\epsilon_{m})].
\end{align*}
This integration vanishes unless the poles of the two Green's functions
locate on the opposite sides of the real axis of the complex plane,
the particle-hole diffusion vertex can be obtained as,

\[
\Lambda(\mathbf{q},i\omega_{n})=\frac{\theta[\epsilon_{m}(\omega_{n}-\epsilon_{m})]}{\tau_{0}\left(|\omega_{n}|+\mathcal{D}\mathbf{q}^{2}\right)},
\]
with the noninteracting diffusion constant $\mathcal{D}=v_{F}^{2}\tau_{0}/3$
and $\tau_{0}$ is the elastic relaxation time induced by disorder
effect. Then the electron-phonon self-energy with vertex correction
from disorder effect {[}shown as Fig. S1(b){]} is given by
\begin{align*}
\Sigma_{ep}^{\prime}(\mathbf{p},i\epsilon_{m}) & =-\frac{1}{\beta}\sum_{i\omega_{n},\mathbf{q}}\frac{|M_{\mathbf{q}}|^{2}D^{(0)}(\mathbf{q},i\omega_{n})\theta(\epsilon_{m}(\omega_{n}-\epsilon_{m}))}{(|\omega_{n}|+\mathcal{D}\mathbf{q}^{2})^{2}\tau_{0}^{2}}\hat{G}^{(0)}(\mathbf{p}-\mathbf{q},i\epsilon_{m}-i\omega_{n})\\
 & =\frac{\Xi^{2}}{\varrho c_{s}^{2}}\frac{1}{\beta V}\sum_{\mathbf{q}}\sum_{\omega_{n}-\epsilon_{m}>0}\frac{1}{(|\omega_{n}|+\mathcal{D}\mathbf{q}^{2})^{2}\tau_{0}^{2}}\hat{G}^{(0)}(\mathbf{p}-\mathbf{q},i\epsilon_{m}-i\omega_{n}),
\end{align*}
where the summation over $\omega_{n}$ is restricted to the region
of $\epsilon_{m}(\omega_{n}-\epsilon_{m})>0$. For $|\mathbf{p}|\sim k_{F}$
and $\epsilon_{m}>0$ ($\epsilon_{m}\to0$), since the dominant contributions
in the integrations are due to both small $q$ and small $|\omega_{n}|$,
$\hat{G}^{(0)}(\mathbf{p}+\mathbf{q},i\epsilon_{m}-i\omega_{n})$
can be approximated by $2\tau_{0}i$ and the $\mathbf{q}$ and $\omega_{n}$
dependences of $|M_{\mathbf{q}}|^{2}D^{(0)}(\mathbf{q},i\omega_{n})\approx\frac{\Xi^{2}}{V\varrho c_{s}^{2}}$
can be ignored,
\[
\Sigma_{ep}^{\prime}(\mathbf{p},i\epsilon_{m})=2\tau_{0}^{-1}i\frac{\Xi^{2}}{\varrho c_{s}^{2}}\frac{1}{\beta V}\sum_{\mathbf{q}}\sum_{\omega_{n}-\epsilon_{m}>0}\frac{1}{(|\omega_{n}|+\mathcal{D}\mathbf{q}^{2})^{2}}
\]
after performing the $\mathbf{q}$ integration, 
\begin{equation}
\Sigma_{ep}^{\prime}(\mathbf{p},i\epsilon_{m})=2\tau_{0}^{-1}i\frac{\Xi^{2}}{\varrho c_{s}^{2}}\frac{1}{8\pi\mathcal{D}^{3/2}}\frac{1}{\beta}\sum_{\omega_{n}-\epsilon_{m}>0}\frac{1}{\sqrt{|\omega_{n}|}},\label{eq:Sigmaep}
\end{equation}
where the summations over $\omega_{n}$ is up to $|\omega_{n}|=\tau_{0}^{-1}$
which corresponds a upper bound of the summation $n_{c}=(2\pi\tau_{0}k_{B}T)^{-1}$
,

\[
\Sigma_{ep}^{\prime}(\mathbf{p},\epsilon)=2\tau_{0}^{-1}i\frac{\Xi^{2}}{\varrho c_{s}^{2}}\frac{\sqrt{k_{B}T}}{4(2\pi\mathcal{D})^{3/2}}\left[\zeta(\frac{1}{2},\frac{\epsilon}{2\pi k_{B}T}+\frac{1}{2})-\zeta(\frac{1}{2},\frac{1}{2\pi\tau_{0}k_{B}T}+1)\right].
\]
At $T=0$ and $|\epsilon|\ll1/\tau_{0}$, from Eq. (\ref{eq:Sigmaep}),
we have
\[
\Sigma_{ep}^{\prime}(\mathbf{p},\epsilon)=\tau_{0}^{-1}i\frac{\Xi^{2}}{\varrho c_{s}^{2}}\frac{1}{8\pi^{2}\mathcal{D}^{3/2}}\int_{\epsilon}^{\tau_{0}^{-1}}dx\frac{1}{\sqrt{x}}\approx i\frac{\Xi^{2}}{\varrho c_{s}^{2}}\frac{3\sqrt{3}}{4\pi^{2}(\hbar v_{F})^{3}}.
\]
For $T\neq0$, and $\epsilon\ll k_{B}T$,
\[
\Sigma_{ep}^{\prime}(\mathbf{p},\epsilon)\approx2\tau_{0}^{-1}i\frac{\Xi^{2}}{\varrho c_{s}^{2}}\frac{\sqrt{k_{B}T}}{4(2\pi\mathcal{D})^{3/2}}\left[\zeta(\frac{1}{2},\frac{\epsilon}{2\pi k_{B}T}+\frac{1}{2})-\zeta(\frac{1}{2},\frac{1}{2\pi\tau_{0}k_{B}T}+1)\right]
\]
where $\zeta(s,a)$ is the Hurwitz zeta function with the asymptotic
expansion for large argument as

\[
\zeta(s,a)\sim\frac{a^{1-s}}{s-1}+\frac{1}{2}a^{-s}+\cdots.
\]
The second term in the blanket gives a constant contribution independent
of $\epsilon$ and $k_{B}T$ which can be absorbed into the disorder
induced self-energy. Let us consider only the first term. For $|\epsilon|\gg k_{B}T$,
we have
\[
\Sigma_{ep}^{\prime}(\mathbf{p},\epsilon)\sim-\tau_{0}^{-1}i\frac{\Xi^{2}}{\varrho c_{s}^{2}}\frac{\sqrt{\epsilon}}{4\pi^{2}(\mathcal{D})^{3/2}},
\]
and for $|\epsilon|\ll k_{B}T$, 
\[
\Sigma_{ep}^{\prime}(\mathbf{p},\epsilon)\sim2\tau_{0}^{-1}i\frac{\Xi^{2}}{\varrho c_{s}^{2}}\frac{\sqrt{k_{B}T}}{4(2\pi\mathcal{D})^{3/2}}\zeta(\frac{1}{2},\frac{1}{2}).
\]
This mixing effect from EPI and the disorder only modifies the imaginary
part of the self-energy when the temperature is very low. In the classical
regime $(T>\Theta_{D})$, this higher order correction becomes negligible.
We believe the lowest order perturbation theory may adequately account
for these observed anomalous transport properties.

\section{Finite temperature conductivity}

After having the self-energy from the EPI, we can calculate the transport
quantities. To distinguish from the quantities of the non-interacting
case, all renormalized quantities such as the effective mass, the
effective chemical potential, and the effective velocity will be denoted
with a bar overhead. In the Kubo-Streda formalism of the linear response
theory \citep{streda1982quantised,Wang18prb}, the conductivity tensor
can be expressed by means of the Green's functions,
\begin{equation}
\sigma_{ij}=\sigma_{ij}^{(1)}+\sigma_{ij}^{(2)}+\sigma_{ij}^{(3)}
\end{equation}
with

\begin{align}
\sigma_{ij}^{(1)}= & \frac{\hbar e^{2}}{4\pi V}\int_{-\infty}^{+\infty}d\epsilon n_{F}^{\prime}(\epsilon-\bar{\mu})\text{Tr}\left[\hat{v}_{i}(G^{R}(\epsilon)-G^{A}(\epsilon))\hat{v}_{j}(G^{R}(\epsilon)-G^{A}(\epsilon))\right]\\
\sigma_{ij}^{(2)}= & -\frac{\hbar e^{2}}{4\pi V}\int_{-\infty}^{+\infty}d\epsilon n_{F}^{\prime}(\epsilon-\bar{\mu})\text{Tr}\left[\left(\hat{v}_{i}G^{R}(\epsilon)v_{j}-\hat{v}_{j}G^{R}(\epsilon)\hat{v}_{i}\right)G^{A}(\epsilon)\right]\\
\sigma_{ij}^{(3)}= & \frac{\hbar e^{2}}{4\pi V}\int_{-\infty}^{+\infty}d\epsilon n_{F}(\epsilon-\bar{\mu})\text{Tr}\Bigg[\left(\hat{v}_{i}G^{R}(\epsilon)\hat{v}_{j}-\hat{v}_{j}G^{R}(\epsilon)\hat{v}_{i}\right)\frac{dG^{R}(\epsilon)}{d\epsilon}\nonumber \\
 & -\left(\hat{v}_{i}G^{A}(\epsilon)\hat{v}_{j}-\hat{v}_{j}G^{A}(\epsilon)\hat{v}_{i}\right)\frac{dG^{A}(\epsilon)}{d\epsilon}\Bigg]
\end{align}
where $\sigma_{ij}^{(1)}$ is symmetric with respect to $i$ and $j$
and contributes to the diagonal elements of the conductivity tensor,
whereas $\sigma_{ij}^{(2)}$ and $\sigma_{ij}^{(3)}$ are antisymmetric
and contribute to the off diagonal elements. $\hat{v}_{i}=\frac{1}{\hbar}\frac{\partial\mathcal{H}_{Dirac}}{\partial k_{i}}$
is the velocity operator for Dirac materials. It is convenient to
work in the basis of the effective Hamiltonian $\mathcal{H}_{eff}(\mathbf{k})=\mathcal{H}_{Dirac}(\mathbf{k})+\mathrm{Re}\Sigma_{ep}^{R}(\boldsymbol{0},0)$.
In this basis, the velocity in $x,y$ direction can be obtained as
$\widetilde{v}_{\perp}^{\zeta}(k)=\frac{1}{\hbar}\frac{\partial\bar{\xi}_{\mathbf{k}}}{\partial k_{x}}$
and the Green's function can be expressed as $\bar{G}_{\zeta}^{R}(\mathbf{k},\epsilon)=[\epsilon-\bar{\xi}_{\mathbf{k}}^{\zeta}-i\mathrm{Im}\Sigma^{R}(\mathbf{k},\bar{\xi}_{\mathbf{k}}^{\zeta})]^{-1}$,
where $\mathrm{Im}\Sigma^{R}(\mathbf{k},\bar{\xi}_{\mathbf{k}}^{\zeta})=\mathrm{Im}\Sigma_{ep}^{R}(\mathbf{k},\bar{\xi}_{\mathbf{k}}^{\zeta})+\mathrm{Im}\Sigma_{imp}^{R}(\mathbf{k},\bar{\xi}_{\mathbf{k}}^{\zeta})$.
Then the relaxation time can be obtained $\tau_{\zeta}(\mathbf{k})=\hbar/[-2\mathrm{Im}\Sigma^{R}(\mathbf{k},\bar{\xi}_{\mathbf{k}}^{\zeta})]$.
$n_{F}^{\prime}(\epsilon-\bar{\mu})\equiv dn_{F}(\epsilon-\bar{\mu})/d\epsilon$
is the energy derivative of the Fermi-Dirac distribution function
$n_{F}(\epsilon-\bar{\mu})$.

In the absence of magnetic field, due to the combined time-reversal
and inversion symmetry of the Dirac Hamiltonian, the off diagonal
components of the conductivity tensor vanish, the longitudinal conductivity
can be evaluated from $\sigma_{ij}^{(1)}=\delta_{ij}\sigma_{D}$\citep{Akkermans2007},

\begin{equation}
\sigma_{D}(T)=\int_{|\bar{m}|}^{\infty}d\omega(-n_{F}^{\prime}(\omega-\bar{\mu}))\sigma_{+}(\omega)+\int_{-\infty}^{-|\bar{m}|}d\omega(-n_{F}^{\prime}(\omega-\bar{\mu}))\sigma_{-}(\omega),
\end{equation}
with the conductivities for the electron ($\zeta=+$) and hole ($\zeta=-$)
carriers, 
\begin{equation}
\sigma_{\zeta}(\omega)=2e^{2}\bar{\nu}_{\zeta}(\omega)\bar{D}_{\zeta}(\omega),
\end{equation}
where the diffusion constants for two types of carriers are defined
as $\bar{D}_{\zeta}(\mu)=\frac{1}{3}\langle(\widetilde{v}_{\perp}^{\zeta})^{2}\rangle_{FS}\bar{\tau}_{\zeta}(\bar{q}_{\omega})$.
In the presence of magnetic field $B$, say, along the z-direction,
the conductivity tensor can be evaluated in the Landau level representation.
Here we are only interested in the semiclassical regime that the self-energy
corrections can be approximated as the zero field results. In this
regime, the transverse conductivity can be calculated as 
\begin{equation}
\sigma_{xx}(B,T)=\int_{|\bar{m}|}^{\infty}d\omega\frac{-n_{F}^{\prime}(\omega-\bar{\mu})\sigma_{+}(\omega)}{[\chi_{+}(\omega)B]^{2}+1}+\int_{-\infty}^{-|\bar{m}|}d\omega\frac{-n_{F}^{\prime}(\omega-\bar{\mu})\sigma_{-}(\omega)}{[\chi_{-}(\omega)B]^{2}+1},
\end{equation}
and the anomalous part of Hall conductivity $\sigma_{ij}^{(3)}$ can
be neglected, the expression of the Hall conductivity in this case
reads
\begin{equation}
\sigma_{xy}(B,T)=\int_{|\bar{m}|}^{\infty}d\omega\frac{-n_{F}^{\prime}(\omega-\bar{\mu})\sigma_{+}(\omega)\chi_{+}(\omega)B}{[\chi_{+}(\omega)B]^{2}+1}+\int_{-\infty}^{-|\bar{m}|}d\omega\frac{-n_{F}^{\prime}(\omega-\bar{\mu})\sigma_{-}(\omega)\chi_{-}(\omega)B}{[\chi_{-}(\omega)B]^{2}+1}
\end{equation}
where $\chi_{\pm}(\omega)=e\bar{\tau}_{\pm}(\omega)/\bar{m}^{*}$
are the mobilities for the conduction and valance bands with the cyclotron
mass $m^{*}=\frac{1}{2}\frac{dq_{\perp}^{2}}{d\mu}$, respectively.
The mobilities for two bands are strongly energy-dependent and can
vary by orders of magnitude. The resistivity can be obtained by inverting
the conductivity tensor. Then, the Hall resistivity $\rho_{xy}=\sigma_{xy}/[\sigma_{xx}^{2}+\sigma_{xy}^{2}]$
and the transverse resistivity $\rho_{xx}=\sigma_{xx}/[\sigma_{xx}^{2}+\sigma_{xy}^{2}]$
. In a weak magnetic field, the Hall resistivity exhibits a linear
dependence with applied magnetic fields. Thus, the Hall coefficient
$R_{H}$ is defined as the ratio of the Hall resistivity and the applied
magnetic field, 
\begin{equation}
R_{H}(T)=\frac{d\rho_{xy}(B,T)}{dB}\Big|_{B=0}.
\end{equation}

The general expressions for the longitudinal and Hall parts of the
thermoelectric coefficients are 
\begin{align}
\alpha_{xx}(B,T) & =\frac{1}{eT}\int_{|\bar{m}|}^{\infty}d\omega\left(\omega-\bar{\mu}\right)\frac{-n_{F}^{\prime}(\omega-\bar{\mu})\sigma_{+}(\omega)}{[\chi_{+}(\omega)B]^{2}+1}+\frac{1}{eT}\int_{-\infty}^{-|\bar{m}|}d\omega\left(\omega-\bar{\mu}\right)\frac{-n_{F}^{\prime}(\omega-\bar{\mu})\sigma_{-}(\omega)}{[\chi_{-}(\omega)B]^{2}+1},\\
\alpha_{xy}(B,T) & =\frac{1}{eT}\int_{|\bar{m}|}^{\infty}d\omega\left(\omega-\bar{\mu}\right)\frac{-n_{F}^{\prime}(\omega-\bar{\mu})\sigma_{+}(\omega)\chi_{+}(\omega)B}{[\chi_{+}(\omega)B]^{2}+1}+\frac{1}{eT}\int_{-\infty}^{-|\bar{m}|}d\omega\left(\omega-\bar{\mu}\right)\frac{-n_{F}^{\prime}(\omega-\bar{\mu})\sigma_{-}(\omega)\chi_{-}(\omega)B}{[\chi_{-}(\omega)B]^{2}+1}.
\end{align}
The Seebeck coefficient $S_{xx}$ and the Nernst signal $S_{xy}$
thus are given by 
\begin{align}
S_{xx} & =\frac{\alpha_{xx}\sigma_{xx}+\alpha_{xy}\sigma_{xy}}{\sigma_{xx}^{2}+\sigma_{xy}^{2}},\\
S_{xy} & =\frac{\alpha_{xy}\sigma_{xx}-\alpha_{xx}\sigma_{xy}}{\sigma_{xx}^{2}+\sigma_{xy}^{2}}.
\end{align}

\end{document}